\documentclass[prb,aps,twocolumn,showpacs,home,floats]{revtex4}
\usepackage{graphics}
\begin{document}

\title{  {\rm\small\hfill (submitted to Phys. Rev. B)}\\
Effect of a humid environment on the surface structure of
RuO${}_2$(110)}

\author{Qiang Sun, Karsten Reuter and Matthias Scheffler}

\affiliation{Fritz-Haber-Institut der Max-Planck-Gesellschaft, Faradayweg 4-6,
D-14195 Berlin, Germany}

\begin{abstract}
Combining density-functional theory and thermodynamics we compute the
phase diagram of surface structures of rutile RuO${}_2$(110) in equilibrium
with water vapor in the complete range of experimentally accessible
gas phase conditions. Through the formation of hydroxyl or water-like
groups, already lowest concentrations of hydrogen in the gas phase are
sufficient to stabilize an oxygen-rich polar oxide termination even at
very low oxygen pressure.
\end{abstract}

\date{Received 29 July 2002}



\maketitle

\section{Introduction}

Metal oxides are employed in a wide range of technological applications,
as for example catalysis (where they may serve as the support or the
active material) and corrosion or wear protection in mechanical systems. 
A microscopic understanding of the function of these compounds requires
knowledge of their often complex surface atomic 
structures,\cite{henrich94,noguera96}
which may be sensitively influenced by temperature and partial pressures
in the surrounding environment. This dependence could be particularly
prominent in oxygen-containing environments, where oxide terminations
of varying stoichiometry are easily perceived as a function of the
abundance of oxygen in the surrounding. The situation becomes even more 
complicated by realizing that also hydrogen is almost always present
in real-life systems, e.g., as a species in the reactant feed gas or 
as a bulk impurity. Then, it is in fact very plausible that surfaces
are terminated by hydroxyl groups, as has recently been exemplified by 
Wang {\em et al.} for the $\alpha$-Al${}_2$O${}_3$(0001) surface.\cite{wang00}

Although it is quite obvious that such variations in the oxide surface
structure and composition may affect the function of these materials in 
applications, systematic studies analyzing this issue are surprisingly
sparse, even for well-defined oxide surfaces.\cite{henrich94,noguera96} 
To this end we describe in the following how the method of {\em
first-principles atomistic thermodynamics} 
\cite{weinert86,scheffler87,kaxiras87,qian88,wang00,wang98,reuter02a,reuter02b}
can be employed to construct a phase diagram of the most stable surface 
structures of an arbitrary oxide surface in equilibrium with its
environment in the complete range of experimentally accessible
temperature and pressure conditions. As specific example we choose the
RuO${}_2$(110) surface in contact with a humid gas phase formed of
hydrogen and oxygen. Recent experiments have not only shown this
surface to be an efficient CO oxidation catalyst \cite{over00,kim01},
but also a sensitive dependence of the measured turnover rate on the
amount of hydrogen present in the system was reported \cite{zang00}. 
In these measurements, RuO${}_2$(110) showed a particularly high 
reactivity for the low temperature oxidation of CO by humid air,
i.e. conditions where most technological catalysts cannot be applied
due to anodic corrosion \cite{zang00}. This renders this oxide an
interesting object of study despite its problematic technological
applicability due to the poisonous nature of Ru.

Similar to the results obtained by Wang {\em et al.} for 
$\alpha$-Al${}_2$O${}_3$(0001) \cite{wang00} we show in the following
that also for RuO${}_2$(110) already lowest amounts of hydrogen in the
gas phase lead to an overwhelming stability of hydrogenated surface
phases. Through the formation of hydroxyl or water-like groups an
oxygen-rich polar oxide termination is then stabilized over the complete
range of accessible oxygen pressures. Concomitantly, the strength of
the O-H bonds formed is so high, that even in ultra-high vacuum (UHV)
annealing temperatures of the order of 600 K will be required to fully
remove hydrogen from this transition-metal (TM) oxide surface. While 
this not only indicates the danger of unintentional hydrogen
contamination in {\em Surface Science} experiments, our findings also 
demonstrate the sensitive dependence of oxide surface structure and
composition on the surrounding gas phase. The latter should be
systematically taken into account when addressing the functionality of
these surfaces in technical applications.

\section{Theory}

In equilibrium with a given environment, characterized by temperature, $T$,
and partial pressures, $\{p_i\}$, the stable surface structure minimizes
the surface free energy, defined as 

\begin{displaymath}
\gamma(T,\{p_i\}) = \frac{1}{A} [ G(T,\{p_i\},N_i) - \sum_i N_i \mu_i(T,p_i) ].
\end{displaymath}

\noindent
Here, $G(T,\{p_i\},N_i)$ is the Gibbs free energy of the finite crystal,
$N_i$ and $\mu_i(T,p_i)$ are number and chemical 
potential of the species of $i$th type, and $\gamma(T,\{p_i\})$ is normalized to 
energy per unit area by dividing through the surface area $A$. In the present 
application to RuO${}_2$(110) in contact with a gas phase containing oxygen and
hydrogen, the surface free energy is thus a function of the three chemical
potentials of Ru, O, and H. Assuming that the surface is also in
equilibrium with the underlying oxide bulk constrains the chemical
potentials of the two oxide components, as their stoichiometric sum has
then to equal the Gibbs free energy of the bulk oxide. As a
consequence the surface free energy can be discussed just in terms of a
dependence on $\mu_{\rm O}$ and $\mu_{\rm H}$, which are both determined by 
the surrounding gas phase reservoirs and thus directly related to 
the experimentally controllable parameters, temperature and partial pressure 
\cite{reuter02a,comment}. 

We also note that not the absolute value of $G(T,\{p_i\},N_i)$ enters the 
calculation of $\gamma(T,\{p_i\})$, but only the difference of the surface 
and the bulk Gibbs free energies. In ref. \onlinecite{reuter02a} we could 
show that for RuO${}_2$(110) in an oxygen environment the vibrational and
entropic contributions to this difference, $\gamma^{\rm vib.}_{\rm RuO_2}(T)$, 
cancel to a large degree. This does, however, not comprise the contributions 
from adsorbate functional groups. Since they have no counterpart in the bulk 
Gibbs free energy, their absolute vibrational energy and entropy enters
into a contribution $\gamma^{\rm vib.}(T)$ to the surface free energy 
of the corresponding phase \cite{reuter02a}. In most cases involving atomic or 
small molecular adsorbates this extra-contribution is fortunately still 
negligible within our aspired accuracy of $\pm 10-15$ meV/{\AA}${}^2$, as there
is only a limited number of (low-energy) vibrational modes that gives rise to 
it. 

\begin{figure}
\scalebox{0.22}{\includegraphics{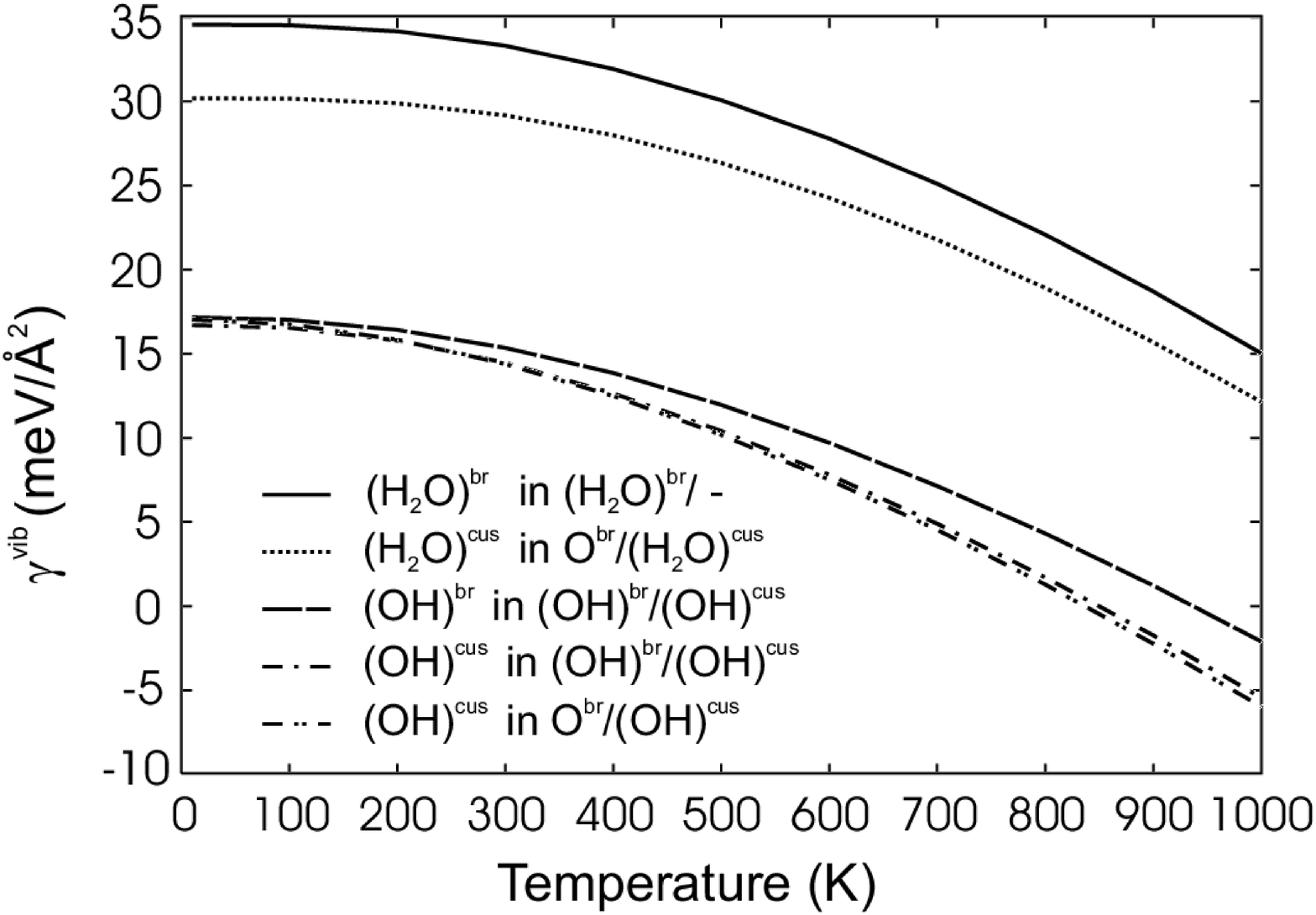}}
\caption{
Vibrational contribution of surface hydrogen groups to the surface 
free energy of various hydrogenated phases of RuO${}_2$(110). See text for
an explanation of the nomenclature used to describe the individual
phases.}
\label{fig0}
\end{figure}

Due to their high-energy modes, hydrogen functional groups form a notable
exception to this line of thought, yielding rather high $\gamma^{\rm 
vib.}(T)$ already for hydroxyl groups or water-like species at
the surface. This is illustrated in Fig. \ref{fig0} for various such phases 
discussed later in the text. The $\gamma^{\rm vib.}(T)$ have been 
obtained from a detailed vibrational analysis of the various surface species,
diagonalizing the complete dynamic matrix while leaving the substrate fixed
\cite{wang03}. We see that for $T < 700$\,K $\gamma^{\rm vib.}(T)$ is of the order of
15 meV/{\AA}${}^2$ for hydroxyl groups and of the order of 30 meV/{\AA}${}^2$
for water-like adspecies. Interestingly, there is little variation in these
values, when the functional groups adsorb in different surface sites, or even
coadsorb closely to each other (see text below for the specific nomenclature
employed to describe the individual surface phases). Very similar $\gamma^{\rm 
vib.}(T)$ are also obtained if we use typical vibrational modes of 
these two surface functional groups when adsorbed on metallic surfaces 
\cite{thiel78}. Although the individual modes change, these variations have 
apparently little effect on the composite quantity $\gamma^{\rm vib.}(T)$. 
Furthermore, if we restrict our discussion to $T <$ 700-800 K, also
the temperature-dependence of this vibrational contribution to the surface free
energy becomes very small, cf. Fig. \ref{fig0}. Correspondingly, we may
approximately take it into account by simply adding a constant 
$\gamma^{\rm vib.}$ of 15 meV/{\AA}${}^2$ per present hydroxyl-group and 30 
meV/{\AA}${}^2$ per present water-like species in the computation of
the $\gamma(T,\{p_i\})$ of corresponding hydrogenated surface phases.

\begin{table}
\caption{\label{tableI}
Experimental and theoretical binding energies, $E_{\rm b}$, as well as
vibration frequencies of isolated H$_2$, O$_2$ and H$_2$O molecules. The
theoretical values are obtained within our DFT-GGA approach
using the same basis set that is also employed in the
slab computations.}

\begin{tabular}{ll | lll}
         &        & $E_{\rm b}$               & $\nu_{\rm stretch}$ (sym/asym)  
& $\nu_{\rm scissor}$ \\ \hline
H${}_2$  & exp    & 4.52 eV/mol. \cite{CRC00} & 546 meV \cite{thiel78}     &  --
- \\
         & theory & 4.50 eV/mol.              & 538 meV                    &  --
- \\
O${}_2$  & exp    & 5.16 eV/mol. \cite{CRC00} & 193 meV \cite{ross72}      &  --
- \\
         & theory & 5.86 eV/mol.              & 180 meV                    &  --
- \\
H${}_2$O & exp    & 9.60 eV/mol. \cite{NBS65} & 454/466 meV \cite{thiel78} & 198 
meV \cite{thiel78} \\

         & theory & 9.81 eV/mol.              & 424/435 meV                & 189 
meV \\
\end{tabular}
\end{table}

Not aiming at an accuracy better than the stated 10-15 meV/{\AA}${}^2$, we may
then replace the two free energies entering $\gamma(T,\{p_i\})$ by their total 
energies, and calculate them using density-functional theory within the full-
potential linear augmented plane wave (FP-LAPW) approach \cite{blaha99} and the 
generalized gradient approximation (GGA) for the exchange-correlation
functional \cite{perdew96}. The high accuracy of the basis set \cite{basis}
employed in these supercell calculations has been detailed in a preceding
publication that discussed the stability of RuO${}_2$(110) in a pure oxygen
environment \cite{reuter02a}. Thus, the numerical accuracy of
differences between $\gamma(T,\{p_i\})$ values of different surface 
structures is better than $\pm 5$ meV/{\AA}${}^2$, while we stress
that the aforedescribed approximate treatment of vibrational and entropic 
contributions does not allow to distinguish between phases which 
$\gamma(T,\{p_i\})$ differ by less than about 10-15 meV/{\AA}${}^2$. 
In comparison to our previous work we were forced to reduce the oxygen
muffin-tin radius to 1 bohr, in order to treat the short O-H bonds in the
present extension to a gas phase containing both oxygen and hydrogen.
Consequently we had to increase the plane wave cutoff in the interstitial
region from 17 Ry to 20.25 Ry for the wavefunctions and from 169 Ry to 400 Ry 
for the potential to maintain the same level of accuracy. Otherwise
the same set-up is used for our computations as detailed before 
\cite{reuter02a}.

To further assess our ability to properly describe particularly the energetics
of O-H bonds within our DFT-GGA approach, we computed the binding
energy, as well as vibration frequencies of free H${}_2$ and H${}_2$O
molecules in the gas phase by diagonalizing their dynamic matrix. 
The results are compiled in Table \ref{tableI} and compared to our
results for molecular O${}_2$. In contrast to the known strong overbinding
in the latter case \cite{perdew96}, the hydrogen-related results
show close agreement with reported experimental values. With a procedure
described in detail in our previous publication \cite{reuter02a}, we
can significantly reduce the large error introduced into the computed
surface free energy due to the oxygen overbinding, giving us finally
good confidence that the chosen set-up is adequate. Checking especially on the 
approximate nature of the exchange-correlation functional we additionally 
repeated the most relevant computations entering into the construction of the 
surface phase diagram within the local-density approximation (LDA) 
\cite{perdew92} and will comment on these findings in the discussion below.

\section{Surface structure of RuO${}_2$(110)}

\begin{figure}
\scalebox{0.5}{\includegraphics{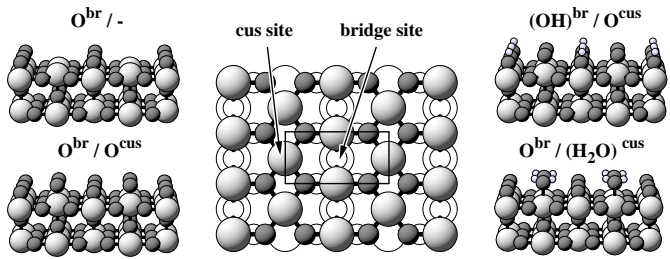}}
\caption{Top view of the RuO${}_2$(110) surface explaining the
location of the two prominent adsorption sites (bridge and cus),
which continue the bulk-stacking sequence (central panel).
Additionally shown are perspective views of the two most stable
$(1 \times 1)$ terminations involving only O (left panel). The
geometry of hydrogenated surfaces is exemplified for two cases,
comprising once a hydroxyl group at the bridge site (top right)
and once adsorbed water at the cus site (bottom right). See text
for explanation of the short hand notation used to describe the
various surface phases. Ru = light, large spheres, O = dark,
medium spheres, H = white, small spheres.}
\label{fig1}
\end{figure}

RuO${}_2$(110) has recently attracted considerable interest as
playing a key role in the catalytic CO oxidation at a Ru(0001)
surface that is exposed to high oxygen pressure.
\cite{reuter02a,reuter02b,over00,kim01,fan01,liu01,wang01,kim01b,boettcher97,boettcher99,boettcher00}. 
The atomic arrangement of this rutile (110) surface, also known from 
TiO${}_2$(110),
exhibits two distinct adsorption sites within the computed (6.43 {\AA}
$\times$ 3.12 {\AA}) rectangular surface unit-cell. Figure \ref{fig1} shows
these two prominent sites over the oxygen-poor termination of
RuO${}_2$(110): The so-called coordinatively unsaturated (cus) site (atop
of fivefold coordinated Ru atoms, Ru${}^{\rm cus}$), and the bridge (br) site
between two fourfold coordinated Ru atoms (Ru${}^{\rm br}$). The bulk
stacking sequence would be continued by oxygen atoms first occupying the
bridge (O${}^{\rm br}$) and then the cus (O${}^{\rm cus}$) sites, thus leading
to the two other $(1 \times 1)$ terminations of a rutile (110) surface shown
in perspective view in the left panels of Fig. \ref{fig1}, cf. 
ref. \onlinecite{reuter02a} for more details.

To evaluate the surface phase diagram we computed what we believe are all
possible $(1 \times 1)$ (co-)adsorption geometries of O and H at the cus
and bridge sites of RuO${}_2$(110). Additionally, we also considered any
possible combination of adsorbed molecular H${}_2$, OH, and H${}_2$O
groups at these two adsorption sites. In the following we will use a
short hand notation to characterize this manifold of studied geometries, 
indicating first the occupancy of the bridge site and then of the cus site, e.g. 
(OH)${}^{\rm br}$/$-$ for an OH-group at the bridge site, the cus site being 
empty, cf. Fig. \ref{fig1}. Although adsorption of H or O is in principle also 
conceivable at other sites on the surface, test calculations always resulted in 
significantly lower binding energies compared to adsorption at the cus or bridge 
sites. Correspondingly, we may restrict our discussion to these two prominent 
sites in the following. In view of the more than 20 studied geometries we would 
like to emphasize at this point that a complete structural relaxation - 
including symmetry breaking at the surface -- was essential to obtain the 
correct energetics, as we find particularly the tilting of OH and water-like 
groups to decisively influence the relative stabilities of the tested 
geometries.

Our self-consistent DFT total energies implicitly contain all lateral
interactions within the considered $(1 \times 1)$ surface unit-cell. 
Significant lateral interactions beyond the first-neighbor cus-br
interaction, cf. Fig. \ref{fig1}, could e.g. lead to ordered adsorption
phases that are more dilute. For pure oxygen phases we therefore performed
a systematic study to extract all lateral interactions between adsorbed
oxygen atoms employing large surface unit-cells up to $(4 \times 1)$ and
$(2 \times 2)$ periodicity \cite{reuter02c}. We found only very small lateral 
interactions, which we understand being a consequence of the rather open surface 
structure (compared to the more often studied close-packed metal surfaces). We 
don't expect this picture to change significantly in the case of H and O 
coadsorption, as confirmed by a series of test calculations employing $(2 \times 
1)$ surface unit-cells. The remaining small lateral interactions or disorder 
effects will only affect the relative stabilities close to the boundaries 
between different stable phases. In the present study we aim to construct the 
large scale phase diagram in the whole range of experimentally accessible gas 
phase parameters. Small variations in the vicinity of the obtained phase 
boundaries do not change the physical conclusions we deduce on the basis of this 
phase diagram and are thus not of present concern. Hence, we will restrict the 
discussion here to the (still large) number of $(1 \times 1)$ geometries and 
defer a systematic study of larger unit-cell phases to a forthcoming 
publication.

\section{Surface phase diagram of hydrogenated RuO${}_2$(110)}

When now discussing the dependence of the computed surface free
energies on the chemical potentials of the gas phase species, we
note that experimentally (and assuming that thermodynamic equilibrium
applies) $\mu_{\rm O}$ cannot be varied without bounds. If it gets too
low, all lattice oxygen atoms would preferably go into the gas phase, i.e
the oxide would decompose. On the other hand, if $\mu_{\rm O}$ gets
too high, oxygen will start to condense on the surface. Theoretical
estimates for both extremes can be derived \cite{reuter02a} and we will
correspondingly restrict the presentation of the surface free energies
to the limited range between this O-poor and O-rich boundary, respectively.

Extending this to a multi-component environment we note that depending
on the gas phase reaction barrier(s) an equilibrium composition may or
may not be present in the gas phase. If the barrier is rather high, 
as in the case of the CO oxidation reaction, the {\em first-principles atomistic 
thermodynamics} concept can e.g. be employed to address the situation
of a {\em constrained equilibrium} where each gas phase species is 
separately in equilibrium with the surface, but not with the
respective other gas component(s). Depending on the reaction barriers
for the adsorbed species, surface kinetic effects may then become
important for certain temperature and pressure conditions, so that
the actual surface structure deviates from the prediction of the 
thermodynamic approach \cite{reuter02b}. 

This problem does not arise, if one concentrates on the situation
where the environment is itself fully equilibrated. In the
present context of a gas phase containing oxygen and hydrogen
this corresponds to allowing for the possibility of water formation in the
oxyhydrogen reaction. In fact, over the whole finite stability range
possible over RuO${}_2$(110), $\mu_{\rm O}$ is already so high 
that water is not only a possibility, but the dominant gas phase 
component, if such a full equilibrium is reached \cite{wang00}. {\em Modelling
this particular situation}, the chemical potential of hydrogen atoms at the 
RuO${}_2$(110) surface is therefore determined by a H${}_2$O reservoir in
the following. To assure that we restrict the discussion to a humid environment, 
i.e. stay within the gaseous state of water for any temperature or pressure, the 
maximum $\mu_{\rm H_2O}$ considered will be at -0.91 eV with respect to the 
total energy of a H${}_2$O molecule, which corresponds to the chemical potential 
of water at the (experimental) critical point.\cite{comment}

\begin{figure}
\scalebox{0.65}{\includegraphics{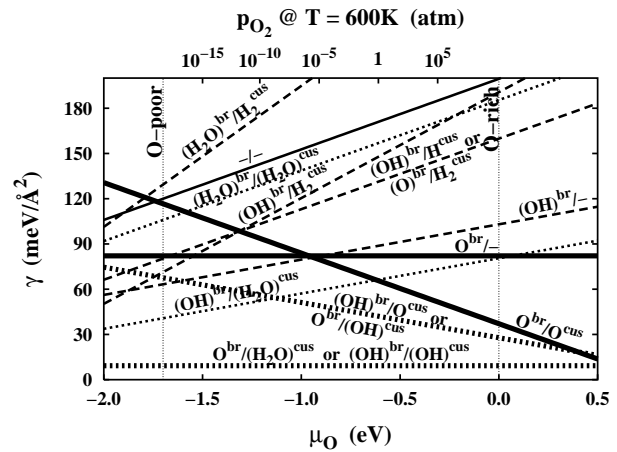}}
\caption{Surface free energies of the three $(1 \times 1)$-RuO${}_2$(110)
terminations involving only oxygen (solid lines). The corresponding
$\gamma(T,\{p_i\})$ of hydrogen containing phases are drawn with dashed
lines, if no O is present at the cus sites, and with dotted lines, if
O is present (see Fig. \ref{fig1} for nomenclature and site explanation).
The dotted vertical lines indicate the allowed range of oxygen chemical 
potential, while the chemical potential of H is in this plot chosen
for very H-rich conditions corresponding to the experimental critical
point of water ($\mu_{\rm H_2O} = -0.91$\,eV \cite{comment}). In the top
$x$-axis, the dependence on $\mu_{\rm O}(T,p_{\rm O_2})$ has been cast into
a pressure scale at a fixed temperature of $T = 600$K.}
\label{fig2}
\end{figure}

Before addressing the equilibrium with the two component gas phase, it is 
instructive to first discuss the oxide surface structure in a pure O${}_2$ 
atmosphere 
\cite{reuter02a}. Then, the surface free energy depends only on the chemical
potential of oxygen, which in turn represents the two-dimensional dependence
on temperature and oxygen pressure. Figure \ref{fig2} shows the limited
stability range between the O-poor and O-rich boundary. To make
the dependence on $\mu_{\rm O}$ a bit more lucid, we additionally
show a pressure scale in the top $x$-axis for fixed $T = 600$\,K, 
which is a typical annealing temperature employed in this system
\cite{over00,kim01}. The solid lines indicate the surface free energies
of the three possible $(1 \times 1)$ terminations of a rutile (110)
surface. While the mixed O-Ru termination, where both cus and bridge
sites are unoccupied (-/-), is rather high up in energy, the two lines 
corresponding to the other two terminations displayed in Fig. \ref{fig2} 
intersect in about the middle of the possible range of oxygen chemical 
potential. At lower $\mu_{\rm O}$, corresponding to ultrahigh vacuum (UHV)
conditions, the traditionally discussed stoichiometric termination is
stable, where oxygen occupies only the bridge sites at the surface
$({\rm O}^{\rm br} / -)$. However, towards higher oxygen pressures
we find a second, O-rich termination to become more stable, in which
O${}^{\rm cus}$ atoms are now also present $({\rm O}^{\rm br} /
{\rm O}^{\rm cus})$.

When next addressing the equilibrium with a gas phase containing additionally
water vapor, the one-dimensional dependence of the surface free energy
on $\mu_{\rm O}$ gets extended to the two-dimensional dependence on
$(\mu_{\rm O}, \mu_{\rm H_2O})$. The maximum effect of the additional
gas component on the surface structure can be seen from the surface
free energies of hydrogenated geometries in the upper H-rich limit of
$\mu_{\rm H_2O}$ described above \cite{comment}. For this particular 
chemical potential ($\mu_{\rm H_2O}$ = -0.91 eV) the lines corresponding to the
most stable structures studied are included in Fig. \ref{fig2}: Dotted lines
represent geometries including O${}^{\rm cus}$ atoms, while dashed lines depict
phases, where no oxygen is present at the cus sites. Interestingly, we find
several cases, where different surface geometries of identical stoichiometry,
which correspondingly exhibit parallel $\gamma(T,\{p_i\})$ lines in Fig.
\ref{fig2}, turn out energetically virtually degenerate, i.e. their surface
free energies within DFT-GGA differ by less than $\pm 5$ meV/{\AA}${}^2$ and are 
thus not distinguishable within the approximations of our approach described 
above. In such cases we only include one line in the graph and label it with 
both geometries.

Particularly relevant for the future discussion will be the two thicker
drawn dotted lines, each representing a degenerate pair of surface phases.
In the first case, two H atoms are present per $(1 \times 1)$ surface
unit-cell, forming either a chemisorbed water-like species at the cus sites 
(${\rm O}^{\rm br} / ({\rm H}_2{\rm O})^{\rm cus}$) or hydroxyl groups
at both bridge and cus sites ($({\rm OH})^{\rm br} / 
({\rm OH})^{\rm cus}$). In the second case, one H atom per unit-cell forms a 
hydroxyl group either at the bridge or at the cus site ($({\rm OH})^{\rm br} / 
{\rm O}^{\rm cus}$ and ${\rm O}^{\rm br} / ({\rm OH})^{\rm cus}$ respectively), 
cf. Fig. \ref{fig1}. We note that particularly the virtual degeneracy of 
these last two phases is somewhat surprising, as they correspond 
to quite different geometries, namely to bridge-bonded and atop-bonded 
OH-groups at different surface sites. As apparent 
from Fig. \ref{fig2}, the surface free energy in this H-rich limit is 
significantly lowered in both cases (thick dotted lines) in comparison to the 
two stable pure O terminations (thick solid lines) discussed before.

\begin{figure}
\scalebox{0.2}{\includegraphics{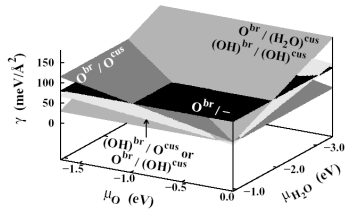}}
\caption{Surface free energies of the four stable phases (or degenerate
phase pairs) discussed in the text, as a function of ($\mu_{\rm O}, \mu_{\rm 
H_2O}$). 
The corresponding phases are drawn as thick lines in Fig. \ref{fig2}
for the H-rich limit.}
\label{fig3}
\end{figure}

Allowing variations of $\mu_{\rm H_2O}$ away from the H-rich limit, the
two-dimensional graph in Fig. \ref{fig2} turns into the three-dimensional
one shown in Fig. \ref{fig3}, where we now only include the surface free
energy planes of the four most stable phases (or degenerate phase pairs),
that will play a role in the later discussion. Note that at the maximum 
$\mu_{\rm H_2O}$ = -0.91 eV along the front $x$-axis in Fig. \ref{fig3} we
simply recover the two-dimensional dependence of Fig. \ref{fig2}. While the 
$\gamma(T,\{p_i\})$ of the two pure O terminations, ${\rm O}^{\rm br} / -$ 
and ${\rm O}^{\rm br} / {\rm O}^{\rm cus}$, do, of course, not depend on 
$\mu_{\rm H_2O}$, the hydrogenated geometries become increasingly less favorable 
the lower the water chemical potential becomes, i.e. the less hydrogen is 
present in the gas phase. For the lowest $\mu_{\rm H_2O}$ shown in Fig. 
\ref{fig3}, all H involving phases have then already become less stable than the 
pure O terminations for any $\mu_{\rm O}$, i.e. the H content in the gas phase 
is so low, that no hydrogen can be stabilized at the surface anymore.

\begin{figure}
\scalebox{0.33}{\includegraphics{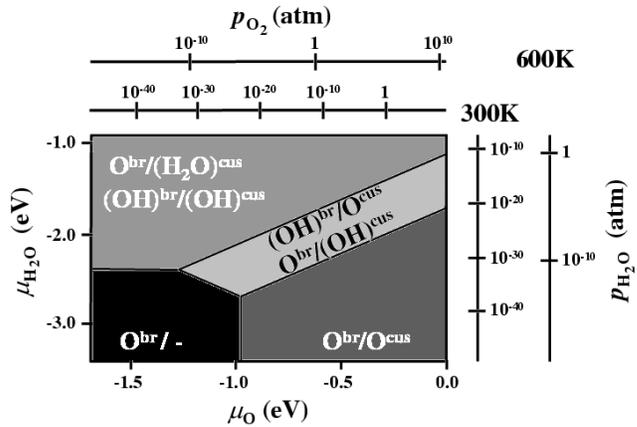}}
\caption{Surface phase diagram of stable structures of RuO${}_2$(110)
in equilibrium with a humid environment, i.e. as a function of
$(\mu_{\rm O}, \mu_{\rm H_2O})$ in the gas phase. The additional
axes show the corresponding pressure scales at $T=300$\,K and $600$\,K.}
\label{fig4}
\end{figure}

Although instructive, the three-dimensional presentation does not allow
an easy access to the most relevant information contained in Fig. \ref{fig3},
namely which phase has the lowest surface free energy for given
gas-phase conditions, i.e. given ($\mu_{\rm O},\mu_{\rm H_2O}$). 
Therefore we display in Fig. \ref{fig4} this surface phase diagram
of stable structures, which corresponds to taking a look at Fig. 
\ref{fig3} from below. We find all four phases (or degenerate phase 
pairs) discussed up to now to be most stable in different 
($\mu_{\rm O}, \mu_{\rm H_2O}$)-regions: At the lowest $\mu_{\rm H_2O}$ shown 
so little hydrogen is present in the environment, that just the surface 
structure dependence of RuO${}_2$(110) in a pure O${}_2$ atmosphere is 
recovered, i.e. the stoichiometric O${}^{\rm br} / -$ phase is stable at low 
$\mu_{\rm O}$, while towards higher oxygen pressures, O${}^{\rm cus}$ is 
additionally stabilized leading to the ${\rm O}^{\rm br} / {\rm O}^{\rm cus}$ 
phase. Upon increasing the H content in the gas phase hydrogen first forms one 
hydroxyl group per surface unit-cell ($({\rm OH})^{\rm br} / {\rm O}^{\rm cus}$
or ${\rm O}^{\rm br} / ({\rm OH})^{\rm cus}$), whereas towards H-rich and
O-poor conditions in the upper left part of Fig. \ref{fig4} the maximum
coverage of two H atoms per surface unit-cell is reached. This leads
either to two hydroxyl groups at both bridge and cus sites ( $({\rm OH})^{\rm 
br} / ({\rm OH})^{\rm cus}$ ) or to a chemisorbed water-like species,
bound with 1eV to the cus sites as shown in the right bottom panel
of Fig. \ref{fig1} (${\rm O}^{\rm br} / ({\rm H}_2{\rm O})^{\rm cus}$).

In order to check how much our results may be affected by the 
exchange-correlation functional, we recomputed the stable surface phases
using the LDA, i.e. considering the LDA lattice constant for the
substrate (6.45 {\AA} $\times$ 3.16 {\AA}) and including a full geometry 
reoptimization. Although the absolute surface free energies are found to be 
on average about $\approx 15$\,meV/{\AA}${}^2$ higher compared to the
DFT-GGA results, the same energetic sequence as displayed in Fig. \ref{fig2} was
obtained. In particular, the two pairs of degenerate surface structures 
discussed before give again surface free energies differing by less than 15 
meV/{\AA}$^2$ from each other and thus exhibit stabilities that cannot be 
distinguished within the current approximations. Correspondingly, the phase 
diagram obtained within the LDA exhibits the same four stable regions and the 
same qualitative structure as the one shown in Fig. \ref{fig4}. Quantitatively,
the phase boundaries of the hydrogenated phases are shifted by about 
0.5\,eV towards lower $\mu_{\rm H_2O}$, indicating that hydrogen bonds 
are stronger within the LDA. In line with our preceding study 
\cite{reuter02a} we therefore conclude that the uncertainty in the DFT
exchange-correlation functional may affect transition temperatures or pressures 
between stable phases deduced on the basis of the constructed phase diagrams. 
Yet the transitions {\em per se}, i.e. the overall structure of the phase 
diagram, are not changed.

To translate the dependence on the chemical potentials into
temperature and pressure, the pressure scales at room temperature and
$T=600$\,K are also included in Fig. \ref{fig4}. From those we can derive 
e.g. what pressures would be required in the gas phase
in order to stabilize hydrogen at RuO${}_2$(110). The transition from the
pure O terminations to the H involving phases at somewhat academic hydrogen 
pressures of the order of $10^{-30}$ atm for $T=300$\,K simply indicates 
that the oxide surface will always be hydrogenated at this low temperature in 
any realistic 
environment, or in other words that the other areas of the phase diagram are
experimentally not accessible at T=300\,K. From our results we can further
conclude, that even under UHV conditions, temperatures of the order of 
500-600 K (GGA, or 600-700 K within the LDA) would be required to
remove hydrogen completely from the surface, while at atmospheric pressure this
rises to about 1000 K, i.e. to temperatures already close to the decomposition
of the whole oxide. It is interesting to notice that in a number of
recent experimental studies on RuO${}_2$(110) an annealing temperature of
only $T = 600$\,K was employed \cite{over00,kim01,kim01b}. 
According to our results this might not have been sufficient to completely
remove residual H contaminants from the surface. Similarly, our results
also imply that hydrogen has to be taken into account as a likely
surface species under all realistic gas phase conditions up to rather
elevated temperatures, e.g. also in all catalytically-relevant
environments.

\section{Comparison with hydrogenated \boldmath $\alpha$-
A\lowercase{l}${}_2$O${}_3$}

The phase diagram in Fig. \ref{fig4} reveals that the most notable 
effect of hydrogenation is to stabilize O${}^{\rm cus}$ atoms over
the complete range of possible oxygen pressure: While in the pure 
O${}_2$ atmosphere the phase including O${}^{\rm cus}$ becomes 
only more stable towards higher oxygen chemical potentials, this 
species is present in all four stable hydrogenated phases for any 
$\mu_{\rm O}$ at sufficiently high $\mu_{\rm H_2O}$, cf. Fig. 
\ref{fig4}. In other words, already lowest concentrations of hydrogen 
in the gas phase are sufficient to stabilize a non-stoichiometric
surface terminated by either hydroxyl groups (one H at O) or 
water-like species (two H at O). This finding is analogous to the results on 
$\alpha$-Al${}_2$O${}_3$(0001) obtained by Wang {\em et al.} 
\cite{wang00}, who reported the purely hydrogen-induced stability of
the O-rich O-plane termination of this corundum structure. Yet, 
while in the latter case the polar termination is never favorable without 
the presence of hydrogen, a humid environment only increases the stability
range of the O-rich surface (in terms of hydroxyl groups) in the case 
of RuO${}_2$(110). 

The hydrogenation goes in both cases hand in hand with
a drastic reduction of the work function. For RuO${}_2$(110) adding
the O${}^{\rm cus}$ atoms first significantly increases the
work function by +1.5\,eV, reflecting the increased dipole
moment of the O-rich ${\rm O}^{\rm br} / {\rm O}^{\rm cus}$ 
termination \cite{reuter02a}. On the contrary, adding hydrogen into
the surface geometry markedly reduces the work function again
by -2.0 eV or -3.5 eV for the phases involving one or two 
hydroxyl groups, respectively. On $\alpha$-Al${}_2$O${}_3$, which was found
to stabilize up to three hydroxyl groups per surface unit-cell, even larger
work function reductions by more than 6\,eV were obtained \cite{wang00}. 

Unfortunately it is not clear, whether the latter 3H per cell
coverage corresponds already to full saturation, as the possibility
of water-like species was not considered in the study by Wang 
{\em et al.}.\cite{wang00} Nevertheless, already for the fully 
hydroxylated $\alpha$-Al${}_2$O${}_3$(0001) surface negative surface free 
energies were obtained in the H-rich limit, indicating the strength of the H-OAl 
bond and consistent with the negative heat of the $\alpha$-Al${}_2$O${}_3$ + 
3H${}_2$O $\rightarrow$ 2Al(OH)${}_3$ reaction \cite{wang00}. On the other hand, 
the weaker H-ORu bond never leads to negative $\gamma(T,\{p_i\})$ even at the 
most H-rich conditions, cf. Fig. \ref{fig2}. This lower bond strength
is also reflected by the result that the fully hydroxylated 
RuO${}_2$(110) surface still exhibits a compression of the topmost
layer distance, although this compression is noticeably reduced
compared to the hydrogen free surface. In contrast it had been found
that the topmost layer is already strongly relaxed outwards at the 
fully hydroxylated corundum Al${}_2$O${}_3$(0001) surface \cite{wang00}.
Finally we note in passing that both studies interestingly obtain a rather 
low site specificity for the formation of OH-groups on the O-rich
surface, as in our case expressed by the degenerate
surface free energies of the $({\rm OH})^{\rm br} / {\rm O}^{\rm cus}$ and 
${\rm O}^{\rm br} / ({\rm OH})^{\rm cus}$ phases. Recalling the large
difference in binding energy between O${}^{\rm br}$ and O${}^{\rm
cus}$,  2.5 eV vs. 1.2 eV with respect to molecular oxygen
\cite{reuter02a,reuter02b}, this result is rather unexpected.

\section{Summary}

In conclusion we presented the {\em ab-initio} based surface phase diagram
of stable structures of RuO${}_2$(110) in equilibrium with a humid
environment formed of oxygen and water vapor. We find the oxide surface
to be completely hydrogenated at room temperature for any realistic gas
phase conditions. Depending on the partial pressures in the gas phase H
is either present in form of hydroxyl groups at bridge and/or cus sites,
or in form of a water-like species at the cus sites of RuO${}_2$(110). A 
complete structural relaxation allowing also for a tilting of the various 
surface groups was found necessary in order to arrive at the correct relative
stabilities of the various phases.

The most notable effect of hydrogen on the oxide surface structure is
the stabilization of a polar, O-enriched surface structure through a
considerable reduction of the finite dipole moment introduced upon
addition of the corresponding terminal oxygen atoms. This finding is
analogous to the one of a previous study on the effect of 
hydrogenation on the stability of the $\alpha$-Al${}_2$O${}_3$(0001) 
corundum surface. The lower H-ORu bond strength does however not lead 
to an instability of the bulk oxide structure, as reflected in the 
reported negative surface free energies of fully hydroxylated 
$\alpha$-Al${}_2$O${}_3$(0001) surfaces under H-rich conditions.
Still, the O-H bonds formed on RuO${}_2$(110) are strong enough that
even in UHV annealing temperatures of the order of $\sim 600$\,K will be
required to completely remove residual H contaminants from the surface.
For atmospheric gas phase pressure this stability extends roughly up
to 1000\,K, indicating the necessity to include H as a likely surface
species when modeling the function of RuO${}_2$ oxide surfaces 
in technological applications like e.g. catalysis.

\section{Acknowledgements}

This work was partially supported by the Deutsche 
Forschungsgemeinschaft (Schwerpunkt ``Katalyse''). Q.S. 
is thankful for an Alexander von Humboldt grant. Valuable
discussions concerning the experimental preparation of hydrogenated
RuO${}_2$(110) with H. Conrad, A. Lobo, K. Jacobi, J. Wang and C.Y. Fan
are gratefully acknowledged.


\begin{thebibliography}{10}
\bibitem{comment}
We use $\mu_{\rm O}(T, p_{\rm O_2}) = \frac{1}{2} \mu_{\rm O_2}(T, p{^0}) +
\frac{1}{2} k_{\rm B}T ln(p_{\rm O_2}/p{^0})$ and $\mu_{\rm H_2O}(T, p_{\rm 
H_2O})
= \mu_{\rm H_2O}(T, p{^0}) + k_{\rm B}T ln(p_{\rm CO}/p{^0})$ with
$p^0 = 1$\,atm and $\mu_{\rm i}(T,p^0)$ taken from ref. \onlinecite{JANAF}.
Experimentally, the critical point of water is at $T=647$\,K,
$p = 22$\,MPa. With these values we obtain $\mu_{\rm H_2O}(647 {\rm K}, 22 {\rm 
MPa})
= -0.91$\,eV, which is considered to represent the H${}_2$O-rich
limit of our study.
\end{thebibliography}
\end{document}